\documentstyle[prb,aps,preprint]{revtex}
\font\msxm msam10\relax
\textfont"E\msxm
\mathchardef\leo"3E2E
\begin{document}
\title{Double Degeneracy and Jahn-Teller Effects in CMR Perovskites}
\author{Jun Zang, A. R. Bishop, H. R\"oder\cite{add} }
\address{
    Theoretical Division and Center for Nonlinear Studies, \\
Los Alamos National Laboratory, Los Alamos, NM 87545}

\maketitle
\draft

\begin{abstract}
{
 Jahn-Teller (JT) electron-phonon
coupling effects in the colossal magnetoresistance
 perovskite compounds $La_{1-x}A_xMnO_3$ are investigated.
Electron-electron correlations between  two degenerate Mn $e_g$
orbitals are studied in the Gutzwiller approximation.
The static  JT distortion and antiadiabatic polaron effects
are studied in a modified Lang-Firsov approximation.
We find that (i) the electron or hole character of the
charge carrier depends on the static JT distortion,
and (ii) due to the two-component nature of the JT coupling,
fluctuations in the JT distortion direction contribute
to the charge transport in similar fashion as the local spins.
}
\end{abstract}

The recently rediscovered ``colossal magnetoresistance'' (CMR)
perovskite-based $La_{1-x}A_xMnO_3$ compounds (${A}=Ca,Sr,Ba$)
exhibit striking magnetic field
and temperature  dependence of the
resistivity \cite{Kus89,Cha93,vonH93,jin94}. Strong correlation between the
electrical resistivity and ferromagnetism in these materials \cite{jonker}
was qualitatively
explained by Zener's double exchange (DE) model \cite{Zen51}.
In this model, the three $t_{2g}$ Mn $d_{xy},d_{xz},d_{yz}$
orbitals are completely filled and the electrons are localized;
thus their spins can be treated as a localized composite
spin $S=3/2$;
the conduction electrons are provided by $e_g$
electrons on the Mn$^{3+}$ ions.
Strong Hund's-rule coupling of strength
$J^H \gg t$ (the hopping matrix element) of the electronically
active $e_g$ electrons to the localized $t_{2g}$ electrons,
results in the motion of electrons spin-polarizing
the localized Mn spins and leading to
ferromagnetism. An applied magnetic field
will polarize the local spins and thus decrease the electrical
resistivity.
The DE model was subsequently refined by
Anderson and Hasegawa\cite{And55}, and de Gennes \cite{deG60}, and
studied more recently \cite{kubo72,fura94,Mil95,jp95,Sar95}
to explain topical experiments. However, it now appears
that the DE model can not adequately explain various experimental
data by itself: Jahn-Teller (JT) electron-phonon coupling
also plays an essential role
 \cite{Mil95,jt-old,ion-cn,ramirez}.

Previously, we studied
a model of combined DE and JT coupling \cite{ours} and found
that the JT coupling can substantially reduce the
resistivity and the Curie temperature
$T_c$. Using this model, we can qualitatively
understand the metal-insulator (MI) transition around
$T_c$: namely, a polaron localized
by the combination of the changing of the
spin polarization cloud around it and
Anderson localization due to
the dramatic increase of spin-disorder scattering near $T_c$.
Here we study the DE model with JT coupling
using {\it two} $e_{g}$ orbitals at each lattice site. One motivation is
to incorporate the doubly-degenerate states to take into
account the mobile carrier density correctly and study the
electron-electron (e-e) on-site correlation effects. A second motivation
is that the doubly-degenerate states act like a pseudospin
which couples to a vector (of two phonon modes). For $T$
close to or above $T_c$, a pseudospin-induced ``double exchange''
effect will then become important (see below).

The Hamiltonian we use consists of the DE part
${\cal H}^{DE}$ and coupling to the lattice represented by ${\cal H}^{JT}$.
${\cal H}^{DE}$ can be derived from the Kondo-lattice Hamiltonian
\begin{eqnarray}
  H^{DE} &=& -t_0 \sum_{\langle i,j\rangle\alpha}  (
  a_{i\alpha}^{\dagger}a_{j\alpha}+b_{i\alpha}^{\dagger}b_{j\alpha}+h.c. ) +
V\sum_i n^a_i n^b_i
\nonumber \\
&-& J_H\sum_i \vec{S}_i\cdot\vec{\sigma}_{\alpha\beta}
(a_{i\alpha}^{\dagger}a_{i\beta}+b_{i\alpha}^{\dagger}b_{i\beta}),
\label{eq:hde0}
\end{eqnarray}
where $\vec{\sigma}$ is the Pauli matrix and $\vec{S}_i$ is the
local spin of the three $t_{2g}$ electrons ($S=3/2$). The operators
$a_{i\alpha}$ ($a_{i\alpha}^{\dagger}$) and
$b_{i\alpha}$ ($b_{i\alpha}^{\dagger}$) annihilate (create) an
electron at the two $e_g$ orbitals ($a$ and $b$) with spin $\alpha$ at
site $i$. The second term in Eq.(\ref{eq:hde0}) describes
the e-e interaction between $a$ and $b$
orbitals on the same site. A detailed
calculation \cite{And55}
showed that for the Hund's-rule coupling $J_H \gg t_0/2$,
the eigenstates of a two-site systems are approximately formed
by electrons all parallel (or all anti-parallel) to the on-site local spins.
Based on the counting of single electronic states, it is natural
to postulate this will also be true for $N$-site systems. So in the limit
of $J_H \gg t_0/2$, ${\cal H}^{DE}$ can be written as
\begin{equation}
  H^{DE}= -t_0 \sum_{\langle i,j\rangle} \gamma_S^{ij} (
  a_{i}^{\dagger}a_{j}+b_{i}^{\dagger}b_{j}+h.c. ) + V\sum_i n^a_i n^b_i ,
\label{eq:hde1}
\end{equation}
where we have omitted the spin index of the electrons since
the electronic spin is fixed by the local spins: the single-electron
states form two bands with a gap $\sim 2J_H \gg k_BT$. Thus
only the lower band is occupied, and it can be assumed there
is no double degeneracy
of each $e_{g}$ orbital.
(This assumption is further strengthened if there
is a large on-site Hubbard interaction).
$\gamma_S^{ij}$ is the bandwidth renormalization due to
the DE mechanism \cite{And55}:
$\gamma_S^{ij} =\left\langle { s_0^{ij} +1/2\over 2S+1} \right\rangle $,
where $s_0^{ij}$ is the total spin of the subsystem
of the two localized spins on sites $i$ and $j$ and the electronic spin.
The angular brackets $<\ldots>$
denote a thermal average.

${\cal H}^{JT}$ can be written as
\begin{eqnarray}
{\cal H}^{JT} &=&\lambda_{JT} \sum_i (Q_{1,i} \tau^z_i + Q_{2,i} \tau^x_i)
\nonumber \\
&+&\sum_i {K\over 2} (Q_{1,i}^2+Q_{2,i}^2)
+\sum_i {M\over 2} (\dot{Q}_{1,i}^2
+\dot{Q}^2_{2,i} ) ,
\end{eqnarray}
where $Q_1=(1/\sqrt{6})(2\delta Z-\delta X-\delta Y)$
and $Q_2=(1/\sqrt{2})(\delta X-\delta Y)$ are
the two local JT distortions \cite{van39,good-book}, with $\delta X^i$
the distance change
between neighboring $O^{2-}$ ions along the $X^i$-axis.
Here we have used pseudospin notation to represent the two degenerate
orbitals: $c_{\bar{\uparrow}}=a$; $c_{\bar{\downarrow}}=b$
with $\bar{\alpha}$ denoting the pseudospin index.
Then $\vec{\tau}_i=\vec{\sigma}c_{i\bar{\alpha}}^{\dagger}c_{i\bar{\beta}}$.
When we define the distortion vector $\vec{Q}$ by
$Q_x = Q_1$ and $Q_z = Q_2$, the JT coupling becomes
$\lambda_{JT} \vec{Q}\cdot \vec{\tau}$. This shows clearly that
the eigenenergy
of the system is independent of the distortion
direction $\vec{Q}/|Q|$. If we
assume that $|Q|$ is fixed,
$\vec{Q}$ can be treated as a local ``classical'' spin.
Then the JT coupling is formally similar to the
electron-spin coupling in Eq.(\ref{eq:hde0}). In the CMR perovskites,
 if the JT electron-lattice coupling is close to
the adiabatic limit and the coupling is strong (i.e. effective
$\lambda_{JT} |Q| \gg t_0/2$), it is thus possible effectively to
have additional
``double exchange''  from the 2-component-doublet JT coupling.
However, there are several mechanisms which will stablize a
specific distortion direction, including
(1) anharmonic lattice interaction \cite{cjt},
(2) higher-order JT interactions \cite{cjt},
(3) interactions between local distortions at different
sites \cite{eng70}. To be specific, we consider only effect (3)
here, by introducing an effective near-neighbor ``ferromagnetic'' coupling.
At low $T$, the JT distortion
will be homogeneous in a certain direction. However, at higher $T$,
fluctuations of the local distortions will induce the
additional ``double exchange'' effects. The local
JT distortion will start to be observed for
 $T \leo T_c$ \cite{ion-cn}. Since $\vec{Q}$ is a classical variable
constrained to the $x-z$ plane instead of a sphere, the fluctuations of
$\vec{Q}/|Q|$ will be larger than a corresponding local spin $\vec{S}$.
Thus its fluctuation is still significant at $T \leo T_c$
even though the Curie temperature $T_c$
is smaller than the ``paramagnetic transition
temperature'' of $\vec{Q}/|Q|$ alone.

The following calculations are restricted to the
metallic phase ($T<T_c$). Incorporating
the above JT distortion-induced ``DE'' effect
introduces an additional bandwidth reduction
factor $\gamma_{JT}$. At low $T$, the distortion
direction is ``frozen'', $\gamma_{JT} \equiv 1$.
At $T$ closer to $T_c$, we will approximate the
stablizing force by a ``ferromagnetic'' coupling between
the $\vec{Q}_i$'s. The strength of this coupling is unknown and
needs to be determined by (e.g. structural) experiments.
After a rotation to make the
distortion direction point along the $z$-axis and
quantization of ${\cal H}^{JT}$,
the Hamiltonian components become
\begin{eqnarray}
{\cal H}^{JT} &=& -\sqrt{\epsilon_p \hbar\omega} \sum_{i\bar{\alpha}}
 n_{i\bar{\alpha}}\sigma^z_{\bar{\alpha}} (B_i^\dagger + B_i)
  +\hbar\omega \sum_i B_i^\dagger B_i ,
\nonumber \\
{\cal H}^{DE} &=& -t_0 \sum_{\langle i,j\rangle\bar{\alpha}}
 \gamma_S^{ij} \gamma_{JT}^{ij}
(  c_{i\bar{\alpha}}^{\dagger}c_{j\bar{\alpha}}+h.c. ) +
V\sum_i n_{i\bar{\uparrow}} n_{i\bar{\downarrow}} .
\label{eq:hde2}
\end{eqnarray}
Here, the electron-phonon coupling strength is
$\epsilon_p = (\lambda_{JT})^2/2K$ , and $\hbar\omega$ is the phonon
frequency.

We treat the dynamical phonons
within the homogeneously modified variational Lang-Firsov approximation
\cite{frwm95} using a
canonical transformation
${\cal{U}}=e^{-S_1(\{\Delta\})} e^{-S_2(\gamma)}$,
where
$S_1=1/2\sqrt{\epsilon_p\hbar\omega}
\sum_i \Delta (B_i^\dagger -B_i) $ is designed to describe
static lattice distortions through the
introduction of a (variational) static displacement field $\Delta$.
$S_2=-\sqrt{\epsilon_p / \hbar\omega}\sum_{i\bar{\alpha}}
\gamma n_{i\bar{\alpha}}\sigma^z_{\bar{\alpha}} (B_i^\dagger -B_i)$
describes the antiadiabatic polaron formation, with $\gamma$ measuring
the degree of the polaron effect.
After the
transformation $\cal{U}$ we can average
$\tilde{\cal H}$ over a squeezed phonon state
$|\Psi_{ph}\rangle = \exp(-\sum_i\tau(B_iB_i-B_i^\dagger B_i^\dagger))
|vac\rangle$ with $\tau$ a variational
parameter \cite{zheng}. The e-e correlations are
approximated by the Gutzwiller approximation \cite{gutz}.
The free energy can then be written as
\begin{eqnarray}
  {\cal F} = -\lambda_a(p_a+d)-\lambda_b(p_b+d)+
(V+2\epsilon_p \gamma(2-\gamma)) d
\nonumber \\
-{1\over \beta}
\sum_{k,i\in\{a,b\}} \ln[1+e^{-\beta(q_i\epsilon_k-\mu+\lambda_i)}]
 -\epsilon_p(2\gamma-\gamma^2)(1-x)
\nonumber \\
+{\hbar\omega \over 4}(\tau^2+\tau^{-2})
+{\Delta^2\over 4\epsilon_p}
+(1-\gamma)\Delta(p_a-p_b)
+{\cal F}_s+{\cal F}_q
\label{eq:fe}
\end{eqnarray}
with the electron dispersion relation given by
$\epsilon_k=2\gamma_s\gamma_{JT}t_0\xi(\gamma)(\cos(k_x)+\cos(k_y)+\cos(k_z))$
with a polaronic band narrowing factor
$\xi(\gamma) = \exp(-\epsilon_p \gamma^2\tau^2/(\hbar\omega))$.
Here we have used the notation
$\bar{\uparrow} \rightarrow a$; $\bar{\downarrow} \rightarrow b$.
$q_i$ are the
discontinuities in the single-particle occupation number
at the Fermi surfaces, which give additional band
narrowing due to e-e interactions: $m_i^*/m_0 \sim 1/q_i$.
The relations between $q_i$ and $p_i$ (probability of occupation
of orbital $i$) and
$d$ (double occupancy) is
\begin{equation}
q_i\equiv { \{[p_i(1-p_a-p_b-d)]^{1/2}+(p_{\bar \imath}d)^{1/2} \}^2
\over (p_i+d)(1-p_i-d) } ,
\label{eq:q}
\end{equation}
where $\bar{a}\equiv b$ and $\bar{b}\equiv a$.
 $x$ is the doping density. Note
the crucial distinction between the static JT
and antiadiabatic polaron distortions:
the double degeneracy of $e_{g}$ is lifted by $(1-\gamma)\Delta$;
in contrast, the antiadiabatic polaron distortion contributes an
effective e-e interaction $2\gamma(2-\gamma)\epsilon_p$.

${\cal F}_s$ and ${\cal F}_q$ in Eq.(\ref{eq:fe})
are free energies from the local
spin $S=3/2$ and JT distortion direction $\vec{Q}/|Q|$, respectively.
We used molecular crystal approximations for both
${\cal F}_s$ \cite{kubo72} and ${\cal F}_q$.
$\; {\cal F}_q$ contributes only at high $T$:
\begin{equation}
{\cal F}_q = l_{eff}\langle Q_z\rangle - \log(\nu_{JT})/\beta
-6J_{eff} \langle Q_z\rangle^2 .
\end{equation}
Here $l_{eff}$ is the effective field for
$\vec{Q}/|Q|$, and
$\nu_{JT} =I_0(\beta l_{eff})$.
$\langle Q_z\rangle = I_1(\beta l_{eff})/I_0(\beta l_{eff})$.
In this molecular crystal approximation,
$\gamma_{JT}=\sinh(2\beta l_{eff})/\left(\pi \beta l_{eff} I_0^2(\beta
l_{eff})\right)$,
where $I_n(x)$ is the $n$-th order modified Bessel function.

The mean-field equations at low $T$ ($\gamma_{JT}\equiv 1$)
are derived by minimizing the free-energy
Eq.(\ref{eq:fe}):
\begin{eqnarray}
\lambda_j - \epsilon_p (1-\gamma)^2(p_j-p_{\bar{\jmath}})
\nonumber \\
=\sum_{i\in\{a,b\}}\int d\epsilon \rho(\epsilon)
({\partial q_i \over \partial p_j})
f(q_i,\lambda_i,\epsilon)
\label{eq:mf0} \\
1-x=\sum_{i\in\{a,b\}}\int d\epsilon \rho(\epsilon)
f(q_i,\lambda_i,\epsilon)
\\
(V+2\epsilon_p \gamma(2-\gamma))
\nonumber \\
=
\sum_{i\in\{a,b\}}\int d\epsilon \rho(\epsilon)
( {\partial q_i \over \partial d}
-{\partial q_i \over \partial p_a}
-{\partial q_i \over \partial p_b})
f(q_i,\lambda_i,\epsilon)
\\
(1-\gamma)[(1-x)-(p_a-p_b)^2]
\nonumber \\
=\gamma {\tau^2 \over \hbar\omega}
\sum_{i\in\{a,b\}}\int d\epsilon \rho(\epsilon)
( q_i \epsilon )
f(q_i,\lambda_i,\epsilon)
\\
(1-{1\over \tau^4})={4\epsilon_p\gamma^2 \over (\hbar\omega)^2}
\sum_{i\in\{a,b\}}\int d\epsilon \rho(\epsilon)
( q_i \epsilon )
f(q_i,\lambda_i,\epsilon)
\\
(h-h_0) {\partial m\over\partial h} \gamma_s
/({\partial \gamma_s\over\partial h})
\nonumber \\
=-\sum_{i\in\{a,b\}}\int d\epsilon \rho(\epsilon)
( q_i \epsilon )
f(q_i,\lambda_i,\epsilon) .
\label{eq:mf}
\end{eqnarray}
Here $f(q_i,\lambda_i,\epsilon)=[1+e^{\beta(q_i\epsilon-\mu+\lambda_j)}]^{-1}$
is the Fermi-distribution function, and $\rho(\epsilon)$ is the
electronic density of states. For the dispersion relation $\epsilon_k
=2t_{eff}(\cos(k_x)+\cos(k_y)+\cos(k_z))$,
$\rho(\epsilon)={1\over 2\pi t_{eff}}\int_0^{\infty} dx
\cos\left({x\epsilon\over 2t_{eff}} \right) [J_0(x)]^3 $, with
$J_0(x)$ the zeroth order Bessel function.
Below, we evaluate
$\Phi(z)=\int_{-\infty}^z dx\rho(x)$, which converges easily
and can be accurately approximated
with high order Chebyshev polynomials.

Eq.(\ref{eq:mf0})-(\ref{eq:mf}) are solved numerically by iteration.
The resistivity is calculated using the same
approximation as in Ref.\cite{kubo72}, which is
comparable to the memory function method \cite{Mil95}.
The resistivity is given by: $\rho=\rho_a\rho_b/(\rho_a+\rho_b)$,
where $\rho_i \propto {(m^*)^2\over k_f^i n_i}
\overline{\Delta (\gamma_s \gamma_{JT})}$
with $m^*\sim 1/t_{eff}$ and
\begin{equation}
\overline{\Delta (\gamma_s \gamma_{JT})}
=\langle ({s_0+1/2\over 2S+1})^2 \rangle\langle \cos^2(\theta_{ij}) \rangle
-(\gamma_s \gamma_{JT})^2
\end{equation}
 with $\gamma_{JT}\equiv 1$ at low $T$:
$\overline{\Delta (\gamma_s \gamma_{JT})}
=\langle ({s_0+1/2\over 2S+1})^2 \rangle -(\gamma_s)^2$.
At higher $T$,  the fluctuation of local JT
distortions begins contributing to the resistivity.
The self-consistent equation for $l_{eff}$ can be
derived easily from minimizing the free energy,
and is similar to Eq.(\ref{eq:mf})
through the substitution: $h \rightarrow l_{eff}$,
$m\rightarrow\langle\cos{\theta \over 2}\rangle$, and
$\gamma_{s} \rightarrow \gamma_{JT}$. The resistivity
is readily calculated given
$\langle \cos^2(\theta) \rangle =
{1\over 2}\left(1+I^2_1(\beta l_{eff})/I^2_0(\beta l_{eff})\right)$.

In Fig.~\ref{fig:hole}, we show the variations of the
electron density of each $e_g$ channel.
It can be seen that the character of the
carrier is strikingly dependent on the {\it static} JT distortion $\Delta$.
For small $\Delta$ (or $\Delta=0$ when there is no JT
coupling), the charge carriers will be electrons,
since the electron density in both channels is lower than
half-filling.
Only with large enough $\Delta$ does the carrier (in the majority
channel) become hole-like. The e-e interaction $V$
can change only the double occupancy but not
the carrier character. (Nagaoka ``ferromagnetism'' is unlikely
here due to the doping $x$ and moderate $V$ interaction
strength). With large JT distortion, because of the creation of
the majority channel, the
e-e interaction $V$ is unimportant for
charge transport.
Based on this result, if experiments (e.g. Hall transport or
thermoelectric power) can detect
the hole-character of the charge carriers, this will
indirectly indicate the presence of a
 large static JT distortion, either local or
homogeneous. It is indeed observed that the thermoelectric power is positive
(hole-like) at $x=0.2$ and negative (electron-like) at $x=0.4$ \cite{Neum95},
which is consistent with Fig.~\ref{fig:hole} in a certain range
of electron-phonon coupling and phonon frequency.
In our choice of model parameters we have used
the hopping matrix element as the energy unit: the temperature
unit is calculated with $t_0=1ev$, so if $t_0=0.2ev$, then
$T=400$ will be $T=80K$. The unit of magnetic field is renormalized
by $H=g\mu_B B$, where $g$ is the effective Land\'e $g$-factor.
Our self-consistent mean-field solution gives $T_c \sim 2550$ for $x=0.2$,
i.e. $T_c \sim 500K$ with $t_0 = 0.2 ev$.

Fig.~\ref{fig:mrh} shows the magnetic-field
dependence of the magnetoresistance ratio $\Delta \rho/\rho_0$
in the low-$T$ ($\gamma_{JT} \equiv 1$) region.
The global features of the $\Delta\rho/\rho_0-H$ curves
are somewhat similar to the experimental data
\cite{Neum95}.
Note that we have not included any additional ``residual''
resistivity mechanisms (disorder, phonon scattering, etc) which
will strongly decrease $\Delta\rho/\rho_0$ at low $T$,
and would need to be subtracted from experimental
data for comparison with our prediction.
 From various experiments,
$\Delta \rho/\rho_0$ can be as large as 0.8 for
moderate to high $T$ around $T_c$, so the spin-fluctuation
scattering considered here will then dominate.
The temperature dependence of the
 magnetoresistivity ratio $\Delta \rho/\rho_0$
is shown in Fig.~\ref{fig:mrt}.
Without the local JT distortion direction
fluctuation ($\gamma_{JT} \equiv 1$), the $\Delta \rho/\rho_0-T$
curve is flat close to $T_c$.
At high $T$ ($T<T_C$ and $\gamma_{JT}\neq 1$),
the effects due to the local JT distortion fluctuations
can make
 $|\Delta \rho/\rho_0|$
increase with $T$, see Fig.~\ref{fig:mrt},
which has also been observed \cite{liu95}. Even without
self-consistently including higher order JT coupling, our calculations
 qualitatively agree with the experiments:
there is a characteristic temperature $T_{p}$ ($T_p \sim 1000$
for the parameters we chose). For $T<T_p$, the local
JT distortion direction is frozen, the resistivity is then due
to fluctuation of local spins and other ``residual''
scattering mechanisms. For $T_p < T <T_c$, the
JT distortion direction fluctuation begins to contribute
to the resistivity, and gives correct magnitude and $T$
dependence of $\Delta \rho/\rho_0$.

In conclusion, we have studied the combined effects
of JT electron-lattice
coupling and the double degeneracy of $e_g$ orbitals
in the CMR perovskites in their metallic phase ($T<T_c$).
We found that without the static JT distortion,
the carriers are electrons. However, with
sufficiently large static JT distortion, the carrier can be hole-like.
Also, the additional scattering effects of the JT distortion (
effective ``double exchange''
effects of the local JT distortion direction) will
be enhanced in the inhomogeneous insulating phase ($T>T_c$) where small
magnetic polarons are localized. This will be reported elsewhere.

{\bf Acknowledgements:} We thank S. Trugman, H.
Fehske, R. Heffner, and D. Rabson for helpful discussions.
Work at Los Alamos
is performed under the auspices of the U.S. DOE.
HR thanks the DFG for providing a travel
grant to visit the Los Alamos National Laboratory.


\begin{figure}[t]
\centerline{
}
\caption{
Occupation density variation of each $e_g$ channel with the
 electron-lattice coupling $\epsilon_p$ at low
$T$.
$t_0 = 1$, $V=6.0$, $\hbar\omega = 0.5$, $T=400$.
$\Delta$ increases monotonically with $\epsilon_p$.
\label{fig:hole} }
\end{figure}

\begin{figure}[t]
\centerline{
}
\caption{
 The variation of $\Delta\rho/\rho_0$ with magnetic field
at low $T$ ($\gamma_{JT} \equiv 1$).
$t_0 = 1$, $\epsilon_p = 2.0$, $V=6.0$, $\hbar\omega = 0.5$.
\label{fig:mrh} }
\end{figure}

\begin{figure}[t]
\centerline{
}
\caption{
The variation of  $\Delta\rho/\rho_0$ with $T$
at  low ($\gamma_{JT}\equiv 1$)
and high $T$ ($T<T_c$) with $J_{eff}=0.1$.
Parameters are the same as in
Fig.2 and $\Delta\rho=\rho(H=0.03)-\rho_0$. Note that
the magnitude of $\Delta\rho/\rho_0$ at low $T$ can not be compared
directly with experiments due to ``residual'' scattering (see text).
\label{fig:mrt} }
\end{figure}
\end{document}